\documentclass[aps,prr,twocolumn,superscriptaddress,amsmath]{revtex4-2}	
\usepackage{graphicx}       
\usepackage{color}
\usepackage[colorlinks,urlcolor=blue,citecolor=blue,linkcolor=blue]{hyperref}
\usepackage{cancel}
\usepackage{orcidlink}
  
\graphicspath{{figures_manuscript/}}
\usepackage{verbatim}  
\usepackage[normalem]{ulem}     
\usepackage{matlab-prettifier}
 
\begin{document}  

\title{Modeling reflection and refraction of freeform surfaces}

\author{J. E. G\'{o}mez-Correa\,
\orcidlink{0000-0002-2399-3661}} 
\email{jgomez@inaoep.mx}
\affiliation{Instituto Nacional de Astrof\'{i}sica, \'{O}ptica y Electr\'{o}nica, Coordinaci\'{o}n de \'{O}ptica, Tonantzintla Puebla 72840, Mexico}

\author{A. L. Padilla-Ortiz\,
\orcidlink{0000-0001-5046-9892}} 
\email{laura.padilla@icat.unam.mx}
\affiliation{CONAHCYT - Instituto de Ciencias Aplicadas y Tecnolog\'{i}a, Universidad Nacional Aut\'{o}noma de M\'{e}xico, Circuito Exterior s/n, Ciudad Universitaria, Coyoac\'{a}n, Ciudad de M\'{e}xico, C.P. 04510, Mexico}

\author{S. Ch\'{a}vez-Cerda\,
\orcidlink{0000-0001-5002-7402}} 
\email{sabino@inaoep.mx}
\affiliation{Instituto Nacional de Astrof\'{i}sica, \'{O}ptica y Electr\'{o}nica, Coordinaci\'{o}n de \'{O}ptica, Tonantzintla Puebla 72840, Mexico}

\date{\today}
\begin{abstract}
In this work, we present a detailed procedure of computer implementation of the laws of refraction and reflection on an arbitrary surface with rotational symmetry with respect to the propagation axis. The goal is to facilitate the understanding and application of these physical principles in a computational context. This enables students and instructors alike to develop simulations and interactive applications that faithfully replicate the behavior of light and sound propagating in a diversity of media separated by arbitrary surfaces. In particular it can help to explore freeform optics. Additionally, we include a practical example demonstrating these implementations using either Matlab or open-source Octave programming language.
\end{abstract}
\maketitle

\section{Introduction} 
The rapid advances in LED technology has opened the necessity to investigate focusing and reflecting properties of variety of surfaces other than the commonly obtained from conics, giving birth to a new technology, freeform Optics \cite{FreeForm}. This technology is based on the Snell's law of refraction and the law of reflection that are fundamental in  optics and acoustics governing what happens to the propagation of light or sound in a medium when they encounter an interface with a different medium.  

\begin{figure}[h!]
\centering\includegraphics[width=\linewidth]
    {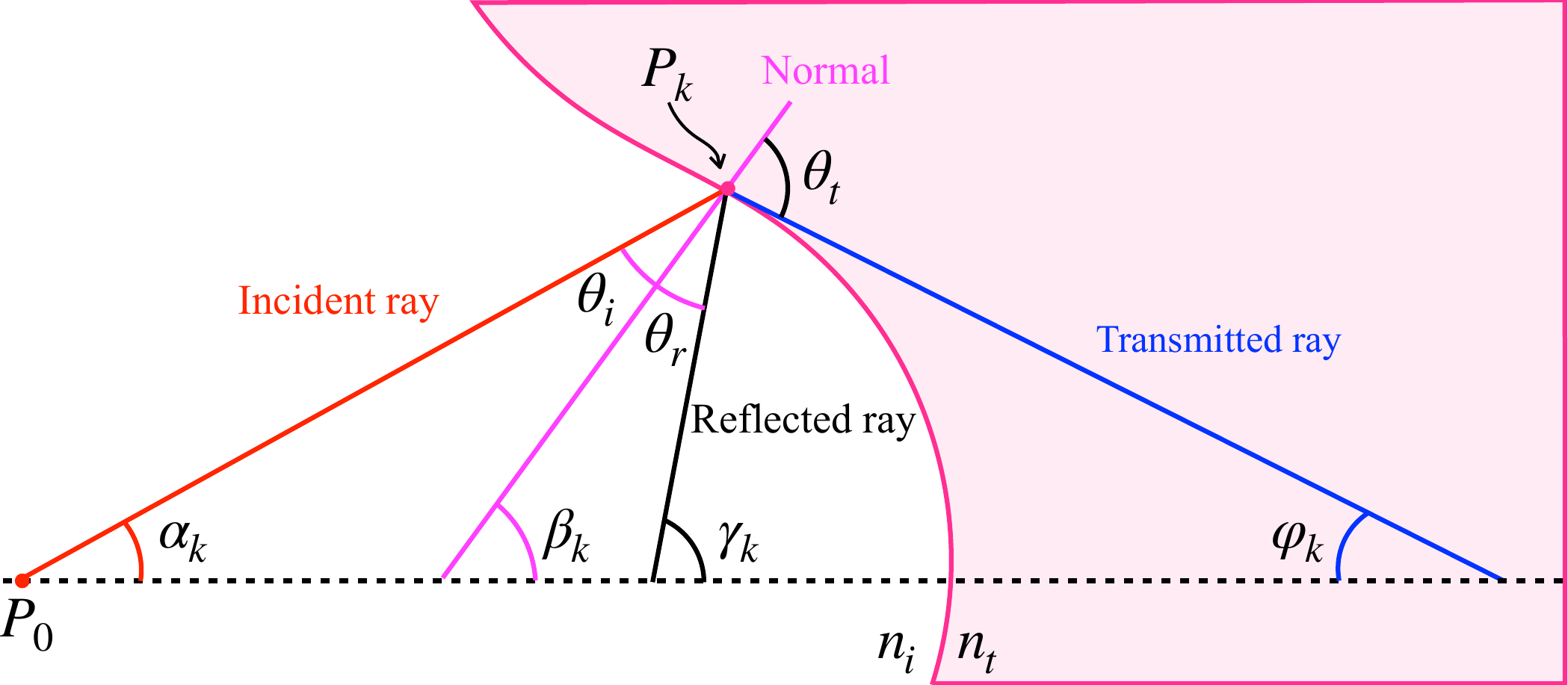}
 \caption{Geometric parameters of incident, reflected, and transmitted rays on a surface.}
 \label{Fig:SnellAndReflectionLaw}
\end{figure}

Refraction Snell's law describes how light and sound change direction when transitioning from one medium to another with different refractive indices, based on their respective angles of incidence and refractive indices \cite{Hecht,Blackstock,Kang_2022}. This principle is essential for understanding image formation in lenses and the propagation of sound in various acoustic environments. Mathematically, this law can be expressed as:
\begin{equation}
n_{i}\sin \theta _{i} =n_{t}\sin \theta _{t}.
\label{SnellLaw}
\end{equation}
Here, a ray, representing the propagation of light or sound, initially propagates in a medium with a refractive index $n_{i}$ and is incident on a medium with a refractive index $n_{t}$. The incident ray forms an angle $\theta_{i}$ with respect to the normal of the surface of the new medium, while the transmitted or refracted ray changes its direction of propagation and travels at an angle $\theta_{t}$ relative to the normal of the surface at the same point, see Fig.\ref{Fig:SnellAndReflectionLaw}.

On the other hand, when light or sound strikes a surface the law of reflection states that the angle of incidence, $\theta_{i}$, equals the angle of reflection, $\theta_{r}$, with respect to the normal to the surface at the point of incidence, see Fig.\ref{Fig:SnellAndReflectionLaw}. In mathematical terms \cite{Hecht,Blackstock}, this law is represented as:
\begin{equation}
\theta _{i}=\theta _{t}.
\label{ReflectionLaw}
\end{equation}
This law is crucial for explaining how acoustic waves reflect in enclosed spaces and how images form in optical mirrors.

Ray tracing is a widely used technique in both optics and acoustics to model wave propagation. In this approach, waves are approximated as rays that propagate in straight lines through homogeneous media and refract or reflect when they encounter interfaces between media with different optical or acoustic properties. In optics, ray tracing is essential in the design of optical systems such as lenses, mirrors, and imaging instruments, where accurately predicting ray paths is crucial for optimizing image quality and reducing aberrations. In acoustics, this technique is useful for predicting sound propagation in enclosed spaces or urban environments, where reflections and refractions from surfaces are key to correctly modeling sound distribution. There are several ray tracing methods, with the most common being: exact ray tracing, paraxial ray tracing, matrix methods, and the $y$-$nu$ method, which uses paraxial ray-trace equations to estimate ray heights and slopes at each surface in an optical system \cite{Hecht,born2013principles,malacara2015optica,ReporteTecnico,Kuttruff,KROKSTAD1968118}.

In this work, we aim to provide readers with the necessary tools to implement Snell's Law of refraction and the law of reflection on an arbitrary surface with rotational symmetry about to the propagation axis, using any programming languaje. In particular we present a practical example providing Matlab code, compatible with Octave programming language (open access software), to demonstrate how these implementations can be executed \cite{MATLAB,Octave,Rogel}. The primary objective is to improve the understanding and extend the application of these physical principles using a computational tool. This will empower students and educators alike to explore and experiment with interactive simulations that accurately depict the behavior of light and sound interacting with diverse freeform structures.

The rest of the paper is organized as follows. Section 2 explains the generation of an arbitrary surface where the rays are incident. Section 3 presents the generation of the incident rays, followed by Section 4, which details the calculation of the normals to the surface at each point where the rays are incident. Section 5 then covers the calculation of the refracted and reflected rays by the surface. In Section 6, a series of classic examples from specialized literature are presented. Section 7 explains the phenomenon of Total Internal Reflection. Finally, the conclusions are provided in Section 8.

\section{Surface of the medium}

In this work, the interface between the two media is described by a curve, as ray tracing is conducted on an arbitrary surface with rotational symmetry around the propagation axis. This means that both the curve and the rays traced along it can be rotated around the propagation axis, generating a three-dimensional ray-tracing model. Therefore, we will explain how such a curve can be generated.

It is well known that a curve in a plane can be represented in three different forms, namely, using explicit functions, implicit functions or parametric functions. Some text also refer to them indistinctly as equations instead of functions \cite{Pre,Lang}. 

An explicit function involves a correspondence rule with one independent variable and one dependent variable, as shown in Eq. (\ref{Expli}):
\begin{equation}
y = f(x),
\label{Expli}
\end{equation}
where $y$ is the dependent variable and $x$ is the independent variable. An example of explicit function is $y=\sqrt{1-x^2}$.

An implicit function, on the other hand, does not allow for a clear distinction between the independent and dependent variables; the dependent variable is not isolated. This is illustrated in Eq. (\ref{Impli}):
\begin{equation}
f(x,y)=a.
\label{Impli}
\end{equation}
The equation of the unit circle, $x^2+y^2=1$, is an example of implicit functions. Notice that one can be tempted to solve for $y$ but then one reaches the point where $y$ is not uniquely determined for a given value of $x$ as it occurs for explicit functions. Another more intricate example is the Mathematician's love equation, $(x^2+y^2-1)^3-x^2y^3=0$, in which it is not possible to solve for any of the variables.

In a parametric function, the variables are written in terms of functions of a third independent variable called a parameter, commonly represented by $t$, and are thus independent of each other, namely,
\begin{eqnarray}
x &=& f(t), \\
y &=& g(t). \nonumber
\label{Parame}
\end{eqnarray}
When the point coordinates $(x,y)$ in the curve are described as functions of $t$, as above, it is said that the curve is parametrized in terms of the parameter $t$.

Generally, to construct a curve, we use the explicit equation or the parametric equation of the desired curve. The implicit equation is rarely used. Due to its simplicity, in this work, we will focus only on explicit and parametric functions.

The first step is to generate the curve computationally once the parametric functions have been established and together with the range of the parameter $t$. The latter consisting of a vector with $m$ elements. We will consider $t_{i}$ and $t_{f}$ as the limits of the desired parameter $t$ and they are such that $x_i=f(t_i)$, $y_i=g(t_i)$ and $x_f=f(t_f)$, $y_f=g(t_f)$ are the end-points of the curve. It is clear that the value of $m$ depends on the desired resolution to obtain smooth graphs, the larger the value the better the resolution. Then we proceed to evaluate the parametric functions.

To illustrate these steps, we will generate a curve using the MATLAB (Octave) programming language. The chosen curve is given by the following parametric equations:

\begin{eqnarray}\label{Paramx}
x(t)&=&a\cos(t)\sin(t)^{2}, \\
y(t)&=&b\sin(t). \nonumber
\end{eqnarray}

The code snippet to generate and plot this curve looks like this:

\begin{lstlisting}[style=Matlab-editor]
% -- Input parametric curve functions --
%
m = 101; %t-parameter sampling
n = 21;  % number of rays
ti = -pi/2.3;
tf = pi/2.3;
t = linspace(ti,tf,m);
a = 3;
b = 3;
x = a.*cos(t).*sin(t).^2;
y = b.*sin(t);
% ------------- Point P_0 --------------
x0 = -10;
y0 = 0;
% ----------- Plot the curve -----------
figure
plot(x,y,'m','linewidth',3)
axis equal
axis([-10 20 -6 6])
set(gca,'fontsize',18,'LineWidth',2)
set(gcf,'Color',[1,1,1])
% --------------------------------------
\end{lstlisting}

\begin{figure}[h!]
\centering\includegraphics[width=\linewidth]
    {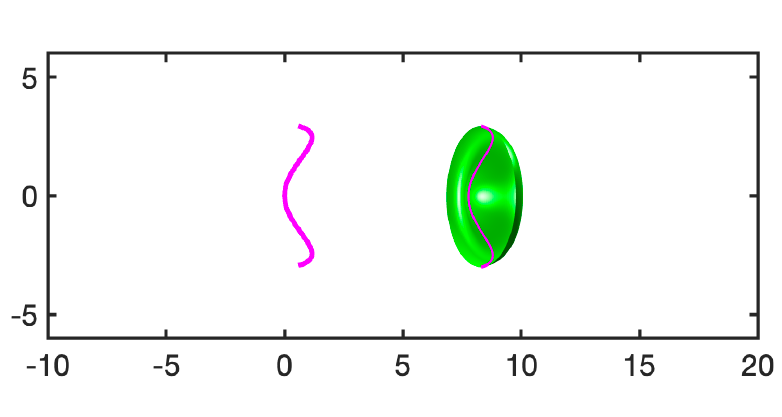}
 \caption{The curve obtained from Eq. (\ref{Paramx}). For visualization purposes, we present the surface generated by rotating the curve around the propagation axis.}
 \label{Fig:Surface}
\end{figure}

This code defines the parameter $t$ over the interval $[-\frac{\pi}{2.3},\frac{\pi}{2.3}]$ with 100 points, calculates the corresponding $x$ and $y$ values using the parametric functions (\ref{Paramx}), and then plots the resulting curve. Fig. \ref{Fig:Surface} shows the plot of the curve described by the given parametric equations. Having generated the curve, we will proceed to create the incident rays.

\section{Incident rays on the curve}

\begin{figure}[h]
 \centering\includegraphics[width=\linewidth]
    {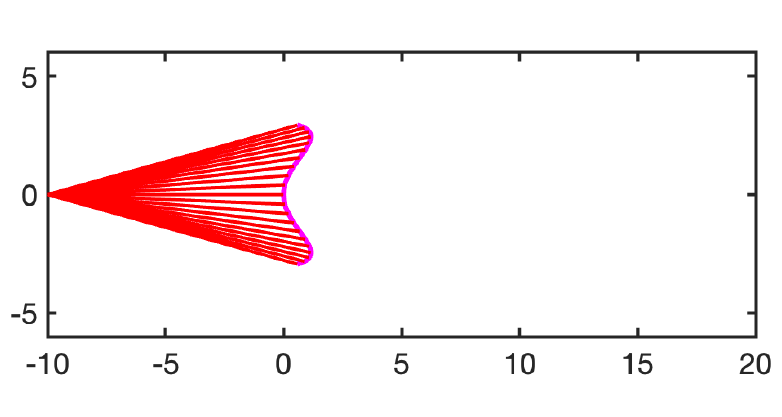}
 \caption{Incident rays on the surface.}
 \label{RayosCurva}
\end{figure}

Let us consider $n$ rays originated from a point source located at $P_0=(x_{0},y_{0})$ arriving at the curve on a point $P_{k}=(x_{k},y_{k})$, with $k=1,2,...n$. Consider $n<m$ to avoid saturation of the plot. The parameter $m$ refers to the number of samples for the parameter $t$, as defined in the code snippet for the input parametric curve functions. Code snippet defining incident points $P_k$:
\begin{lstlisting}[style=Matlab-editor]
% -- Rays from P_0 to Curve at P_{k} --
% Parameter t is redefined to n elements
% Coordinates (x(k),y(k)) are calculated
% ---------- Curve points P_k ----------
t = linspace(ti,tf,n);
x = a.*cos(t).*sin(t).^2;
y = b.*sin(t);
% --------------------------------------
\end{lstlisting}
Since the points $P_k$ on the curve are given by the parametric functions (\ref{Paramx}), to plot each of the rays emanating from the point source $P_0$ we use the equation of the straight line in two point form, namely
\begin{equation}
y=\frac{y_{k}-y_{0}}{x_{k}-x_{0}}\left(x-x_{0}\right)-y_{0}.
\label{Ec_Recta}
\end{equation}
Next is the code snippet to implement these equations:
\begin{lstlisting}[style=Matlab-editor]
% Incident Rays plotted from P_0 to P_k
hold on; %Keep curve to plot Rays
for k = 1:n
     xi = linspace(x0,x(k),m);
     mi = (y(k)-y0)/(x(k)-x0);
     yi = mi*(xi-x0)-y0;
     plot(xi,yi,'r','LineWidth',1);
end
hold off; %Release plot
axis equal
axis([-10 20 -6 6])
set(gca,'fontsize',18,'LineWidth',2)
set(gcf,'Color',[1,1,1])
clear xi yi mi
% --------------------------------------
\end{lstlisting}
The results are shown in Fig. \ref{RayosCurva}.

Once having defined the points where the incident rays intersect the surface, the next step involves calculating the surface normal at each of these points.

\section{Normal lines}

A normal line to a curve at a specific point $P_{n}$ is a line that is perpendicular to the tangent of the curve at that point. The normal line intersects the curve at the point of tangency and has a slope that is the negative reciprocal of the slope of the tangent at that point. The equation of the normal line can be expressed as:
\begin{equation}
y_{N}= -\frac{1}{M_{k}}(x_{N}-x_{k})+y_{k},
\label{NormalEq}
\end{equation}
where $(x_{k},y_{k})$ is the point of tangency, i.e., the points $P_{k}$ and $M_{k}$ is the slope of the tangent to the curve at that point.

For a surface defined by a parametric function, the slope of this tangent line is given by \cite{Pre,Lang}
\begin{equation}
M_{k}=\left.\frac{dy}{dx}\right|_{P_{k}}=\frac{y'(t_{k})}{x'(t_{k})},
\label{NormalDer}
\end{equation}
where $y'(t_{k})=\left.\frac{\mathrm{d}y(t)}{\mathrm{d}t}\right|_{P_{k}}$, and $x'(t_{k})=\left.\frac{\mathrm{d}x(t)}{\mathrm{d}t}\right|_{P_{k}}$. This expression represents the derivative of $y$ with respect to $x$ evaluated at the point \( P_k \) in terms of the derivatives with respect to the parameter $t$.

Notice that if we substitute Eq. (\ref{NormalDer}) into Eq. (\ref{NormalEq}), we obtain the equation of the normal line at each of the points $P_{k}$ on the surface.

For our example, the derivatives of the parametric equations $x(t)$ and $y(t)$ are:\\
\\
For $x(t)=a\cos(t)\sin^2(t)$:
\begin{equation}
\left.\frac{dx}{dt}\right|_{P_{k}}=a\left[2\cos^2(t_{k})\sin(t_{k})-\sin^3(t_{k})\right].
\end{equation}
For $y(t)=b\sin(t)$:
\begin{equation}
\left.\frac{dy}{dt}\right|_{P_{k}}=b\cos(t_{k}).
\end{equation}
Then, the slope of the tangent line at the point $P_k$ in the curve is:
\begin{equation}
M_{k}=\frac{b\cos(t_{k})}{a\left[2\cos^2(t_{k})\sin(t_{k})-\sin^3(t_{k})\right]},
\end{equation}
with this equation, we can determine the equation of the normal line at each of the points $P_{k}$ on the curve, which is given by:
\begin{equation}
y_{N}= \frac{a\left[\sin^3(t_{k})-2\cos^2(t_{k})\sin(t_{k})\right]}{{b\cos(t_{k})}}(x_{N}-x_{k})+y_{k}.
\end{equation}

\begin{figure}[h!]
\centering
\includegraphics[width=\linewidth]{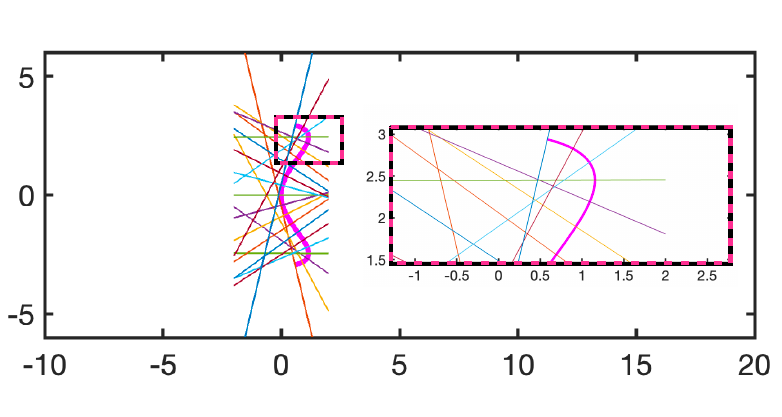}
\caption{Normals to the surface (curve).}
\label{NormalesALaCurva}
\end{figure}

To implement this equation we define a domain for $x_N$ large enough containg the $x_k$ of the curve; the code snippet in MATLAB is as follows:
\begin{lstlisting}[style=Matlab-editor]
% ----------- Normal lines -------------
% ---- x-interval for normals [x1,x2]---
x1 = -2;
x2 = 2;
hold on; %Keep curve to plot Normals
for k = 1:n
    xN = linspace(x1,x2,m);
    Mn = (b*cos(t(k)))/(a*(2*cos(t(k))^2*sin(t(k))-sin(t(k))^3));
    yN = -(1/Mn)*(xN-x(k))+y(k);
    plot(xN,yN,'LineWidth',1);
end
hold off; % Release plot
axis equal
axis([-10 20 -6 6])
set(gca,'fontsize',18,'LineWidth',2)
set(gcf,'Color',[1,1,1])
% --------------------------------------
\end{lstlisting}
Notice that the code finds the normal lines to the curve at each point $P_{k}$ as shown in Fig. \ref{NormalesALaCurva}.
Once the lines normal to the curve have been calculated, we will proceed to calculate the reflected and transmitted rays.

\section{Reflected or transmitted rays by the surface}
With all of the above, we have the necessary to obtain the refracted an reflected trajectories of the incident light or sound at the surface according to the Snell's and reflection laws described in the introduction.  For this purpose, we endeavor to determine the slopes of the reflected and transmitted rays. We will make use of auxiliary angles $\alpha_k$, $\beta_k$, $\gamma_k$ and $\varphi_k$ defined with respect to the Cartesian reference frame and determined by the point $P_k$ at the curve as shown in Fig. \ref{Fig:SnellAndReflectionLaw}.

\begin{figure}[h!]
\centering
\includegraphics[width=\linewidth]{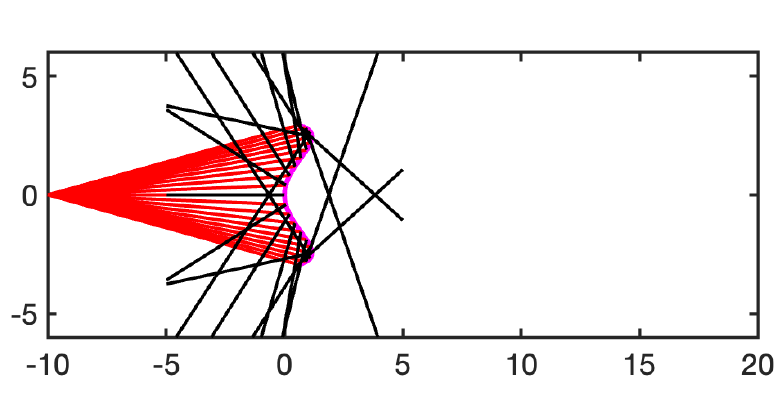}
\caption{Rays reflected by the surface.}
\label{Fig:Reflejados}
\end{figure}

A simple trigonometric calculation shows that the auxiliary reflected angle is given by $\gamma_{k}=2\beta_{k}-\alpha_{k}$ with $\beta_{k}=\tan^{-1}\left(-\frac{1}{M_{k}}\right)$, $\alpha_{k}=\tan^{-1}\left(m_{k}\right)$; $m_k$ and $M_k$ the slopes of the incident ray and of the tangent to the curve at point $P_k$, respectively. Then, the slope of the reflected ray at any point $P_k$ at the curve is given by
$m_{rk}=\tan(\gamma_{k})$ and the reflected rays are determined by the line equation given by
\begin{equation}
y_{rk}=\tan\left(\gamma_{k}\right)\left(x_{rk}-x_{k}\right)+y_{k}.
\end{equation}
Below is the corresponding Matlab (Octave) code snippet to plot the reflected rays shown in Fig. \ref{Fig:Reflejados}.

\begin{lstlisting}[style=Matlab-editor]
% ----------- Reflected rays -----------
hold on; %Keep curve to plot Normals
for k = 1:n
    Mk = (b*cos(t(k)))/(a*(2*cos(t(k))^2*sin(t(k))-sin(t(k))^3));
    betak = atan(-1/Mk);
    mk = (y(k)-y0)/(x(k)-x0);
    alphak = atan(mk);
    gammak = 2*betak-alphak;
    gammakGrad = gammak*180/pi;
    if (abs(gammakGrad)>90)
        xEnd = 6;
    else
        xEnd = -6;
    end
    xrk = linspace(x(k),xEnd,m);
    yrk = tan(gammak)*(xrk-x(k))+y(k);
    plot(xrk,yrk,'k','LineWidth',1);
 end
hold off; % Release plot
axis equal
axis([-10 20 -6 6])
set(gca,'fontsize',18,'LineWidth',2)
set(gcf,'Color',[1,1,1])
% --------------------------------------
\end{lstlisting}

It is important to note that the reflected rays are plotted from the points $P_{k}$, where the rays strike the curve, to a certain plane located either in front of or behind the curve. The latter case occurs for some curves or physical situations in which the reflected rays propagate towards another point at the curve, as can be observed in Fig. \ref{Fig:Reflejados} near the endpoints of the curve. At this point, for simplicity these secondary reflections will be neglected, but if required their trajectories can be obtained following the procedure just described. The choice of plane depends on the value of the slope: if $\gamma_{k}<90^{\circ}$, a plane located before the curve is chosen; if $\gamma_{k}>90^{\circ}$, a plane located after the curve is chosen.

\begin{figure}[h!]
\centering
\includegraphics[width=\linewidth]{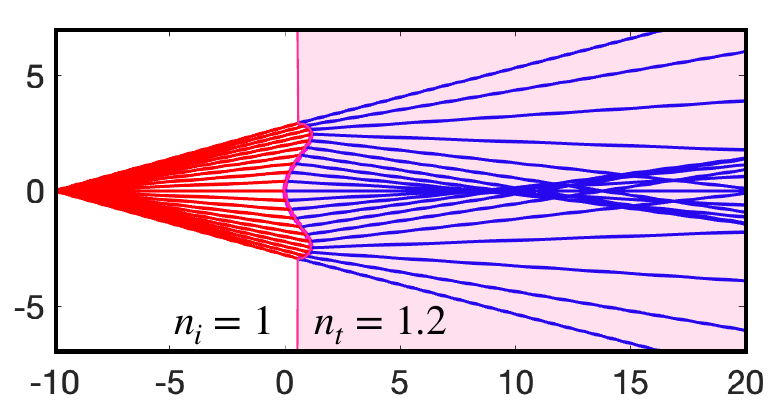}
\caption{Rays transmitted through the surface.}
\label{Fig:Refractados}
\end{figure}

For the transmitted rays, the slope is given by (see Fig \ref{Fig:SnellAndReflectionLaw}):
\begin{equation}
m_{tk}=\tan\left(\varphi_{k}\right),
\end{equation}
where
\begin{equation}
\varphi_{k}=\beta_{k}-\theta_{t}.
\end{equation}
The value of $\theta_{t}$ is easily found using Snell's law, that is,
\begin{equation}
\theta_{t}=\sin^{-1}\left(\frac{n_{1}}{n_{2}}\sin\theta_{i}\right).
\label{Eq:ThetaT}
\end{equation}
Then, the equation of the line with which we will plot the refracted rays is given by:
\begin{equation}
y_{tk}=\tan\left(\varphi_{k}\right)\left(x_{tk}-x_{k}\right)+y_{k}.
\end{equation}
The transmitted rays will be plotted from the curve to a plane located after the curve, as shown in the following Matlab code snippet and in Fig. \ref{Fig:Refractados}.

\begin{lstlisting}[style=Matlab-editor]
% ---- Transmitted or refracted rays ---
% ------ Define refractive indexes -----
n1 = 1;
n2 = 1.2;
hold on; %Keep curve to plot Normals
for k = 1:n
    Mk = (b*cos(t(k)))/(a*(2*cos(t(k))^2*sin(t(k))-sin(t(k))^3));
    betak = atan(-1/Mk);
    mk = (y(k)-y0)/(x(k)-x0);
    alphak = atan(mk);
    gammak = betak-alphak;
    Thetat = asin((n1/n2).*sin(gammak));
    xtk = linspace(x(k),20,m);
    ytk = tan(gammak-Thetat)*(xtk-x(k))+y(k);
    plot(xtk,ytk,'b','LineWidth',1);
end
hold off; % Release plot
axis equal
axis([-10 20 -6 6])
set(gca,'fontsize',18,'LineWidth',2)
set(gcf,'Color',[1,1,1])
% --------------------------------------
\end{lstlisting}
 
Observe that in this example, the rays are incident from a medium with a refractive index of $n_{1} = 1$ into a medium with a refractive index of $n_{2} = 1.2$.

\begin{figure}[h!]
\centering
\includegraphics[width=\linewidth]{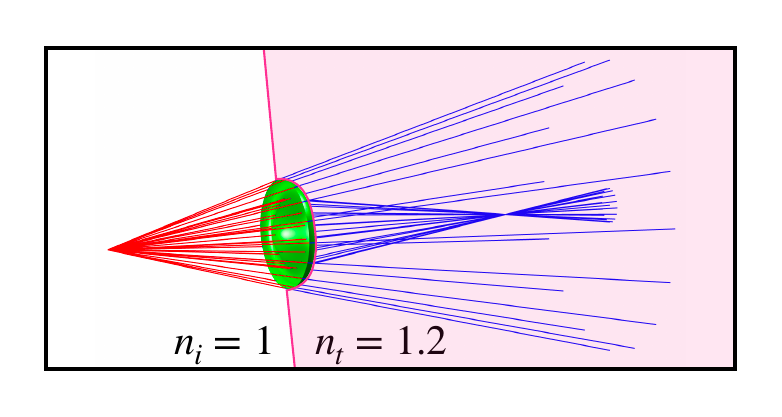}
\caption{Three-dimensional ray tracing through the surface.}
\label{Fig:3D}
\end{figure}

Only for visualization purposes, in Fig. \ref{Fig:3D} we show the ray tracing on the three-dimensional surface generated using Eq. (\ref{Paramx}). This ray tracing was performed by applying a three-dimensional rotational matrix to rotate the entire system from Fig. \ref{Fig:Refractados}.

\section{Other examples}

\begin{figure}[h!]
\centering
\includegraphics[width=0.6\linewidth]{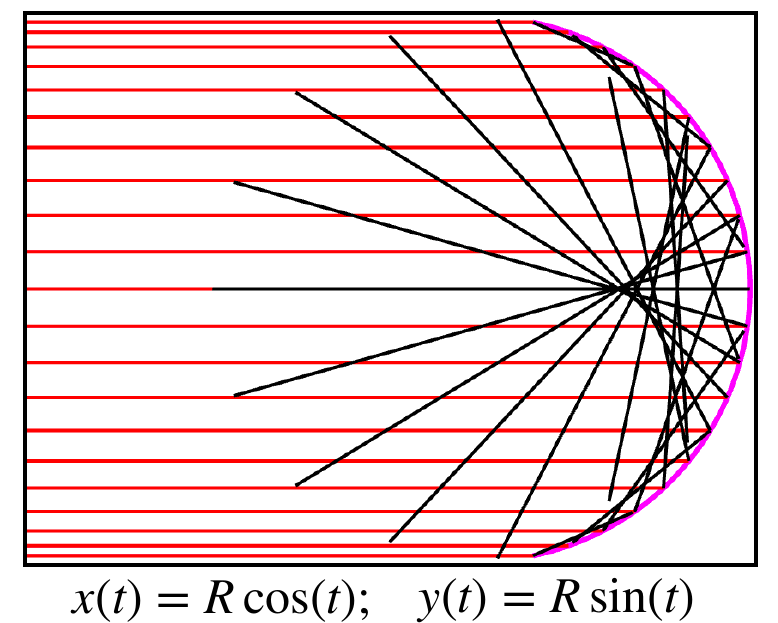}
\caption{Rays reflected by a circular mirror.}
\label{Fig:CircularR}
\end{figure}

\begin{figure}[h!]
\centering
\includegraphics[width=0.6\linewidth]{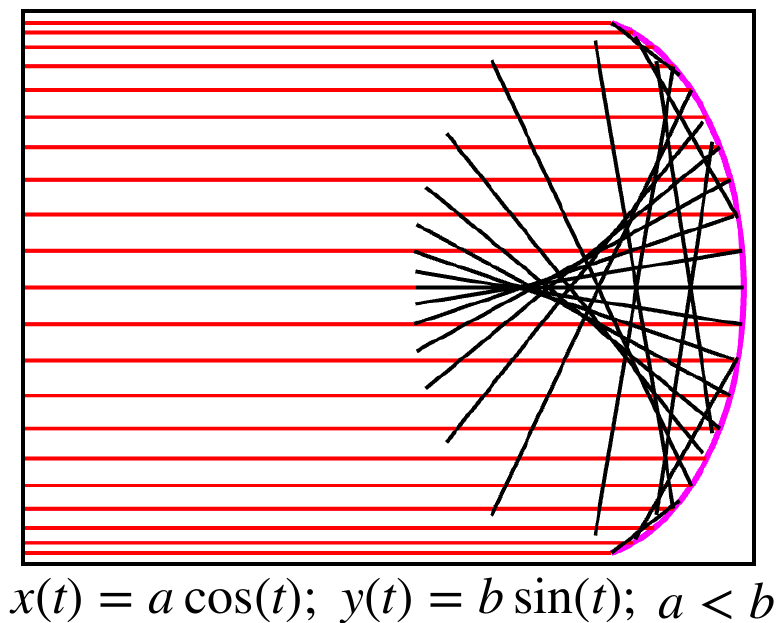}
\caption{Rays reflected by an elliptical mirror with $a<b$ (prolate surface).}
\label{Fig:ElipticoProlateR}
\end{figure}

\begin{figure}[h!]
\centering
\includegraphics[width=0.6\linewidth]{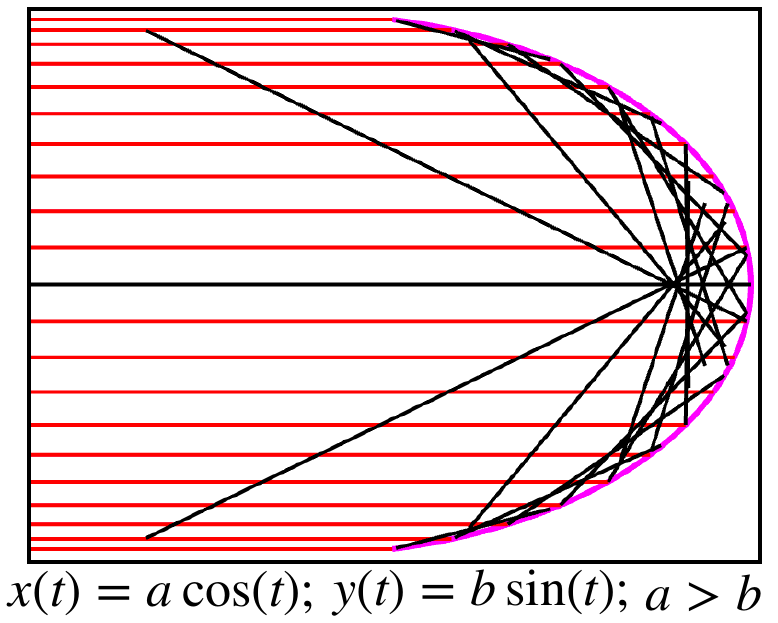}
\caption{Rays reflected by an elliptical mirror with $a>b$ (oblate surface).}
\label{Fig:ElipticoOblateR}
\end{figure}

\begin{figure}[h!]
\centering
\includegraphics[width=0.6\linewidth]{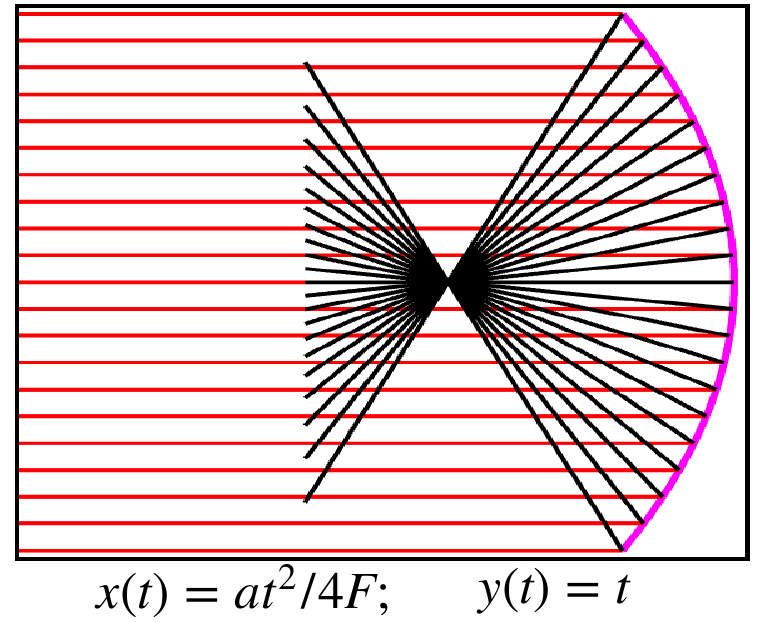}
\caption{Rays reflected by a parabolic mirror. $F$ is the focus of the parabola.}
\label{Fig:ParabolicR}
\end{figure}

\begin{figure}[h!]
\centering
\includegraphics[width=0.6\linewidth]{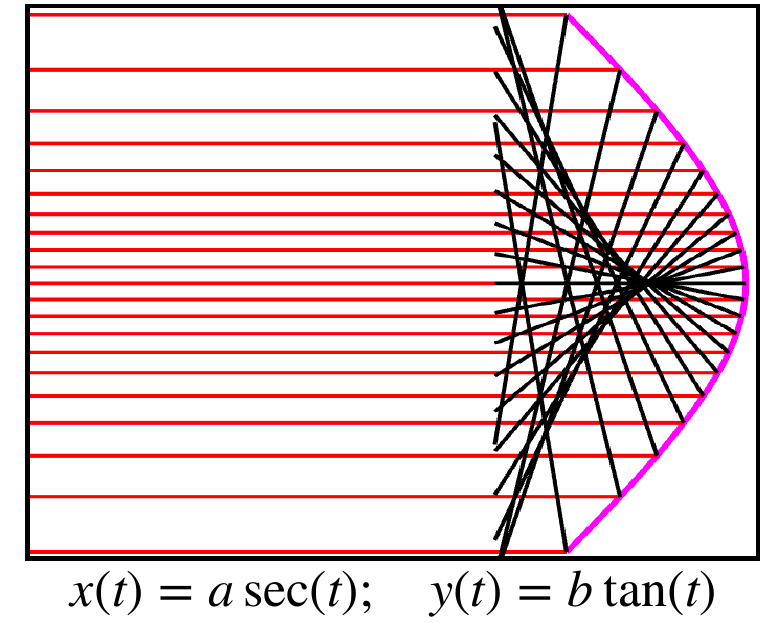}
\caption{Rays reflected by a hyperbolic mirror.}
\label{Fig:HyperbolicR}
\end{figure}

We have developed above a simple code capable of calculating the reflected and refracted rays from an arbitrary curve given by the parametric functions Eq. (\ref{Paramx}). However, the code is general and it can work for any other curve expressed in its parametric form. The only required modification is defining the parametric functions for the coordinates of the curve in question. For example, we can calculate the reflected rays by a circular, elliptical, parabolic, or hyperbolic mirror, as shown in Figs. \ref{Fig:CircularR}, \ref{Fig:ElipticoProlateR}, \ref{Fig:ElipticoOblateR}, \ref{Fig:ParabolicR}, and \ref{Fig:HyperbolicR}, respectively. 

\begin{figure}[h!]
\centering
\includegraphics[width=0.65\linewidth]{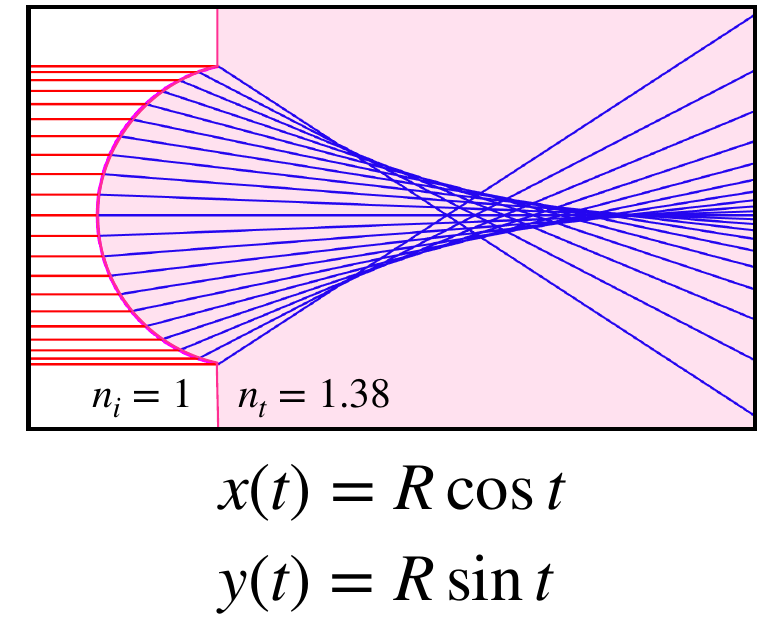}
\caption{Rays refracted by a circular surface.}
\label{Fig:CircularT}
\end{figure}

\begin{figure}[h!]
\centering
\includegraphics[width=0.65\linewidth]{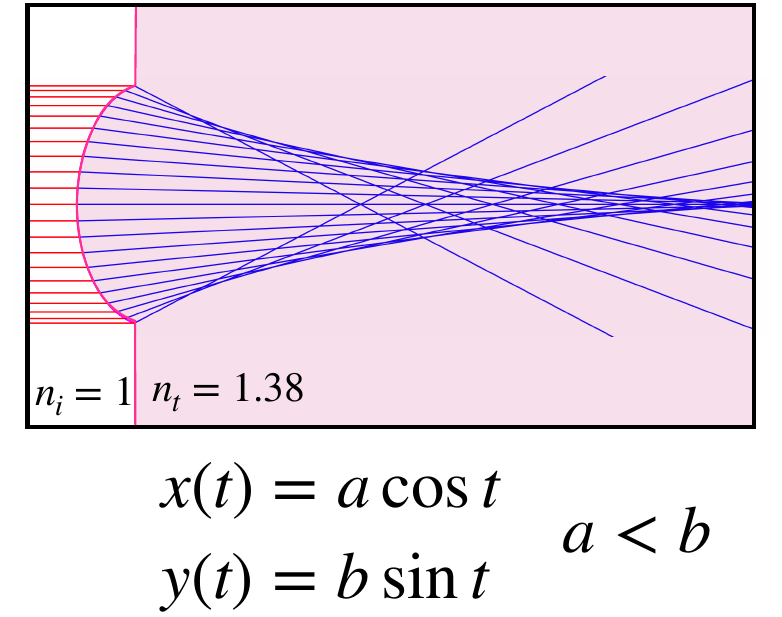}
\caption{Rays refracted by an elliptical surface with $a<b$ (prolate surface)}
\label{Fig:ElipticoProlateT}
\end{figure}

\begin{figure}[h!]
\centering
\includegraphics[width=0.65\linewidth]{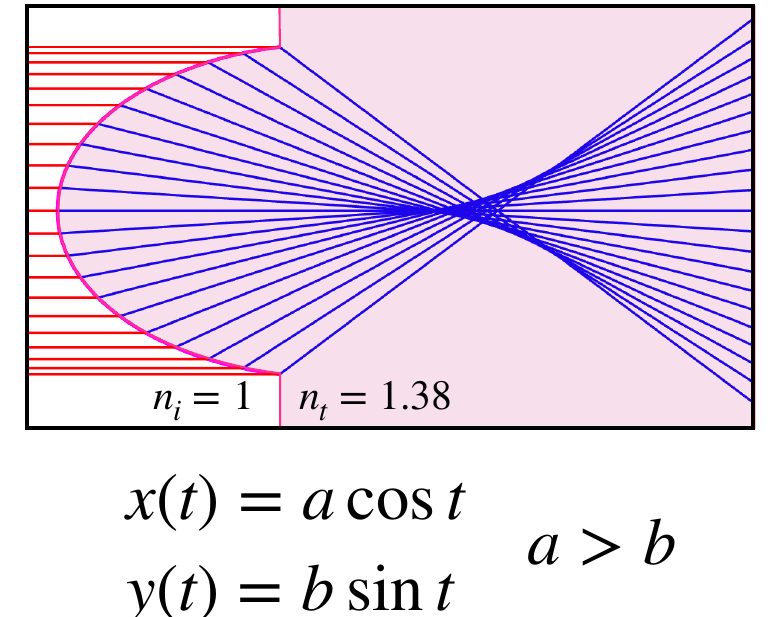}
\caption{Rays refracted by an elliptical surface with $a>b$ (oblate surface).}
\label{Fig:ElipticoOblateT}
\end{figure}

\begin{figure}[h!]
\centering
\includegraphics[width=0.65\linewidth]{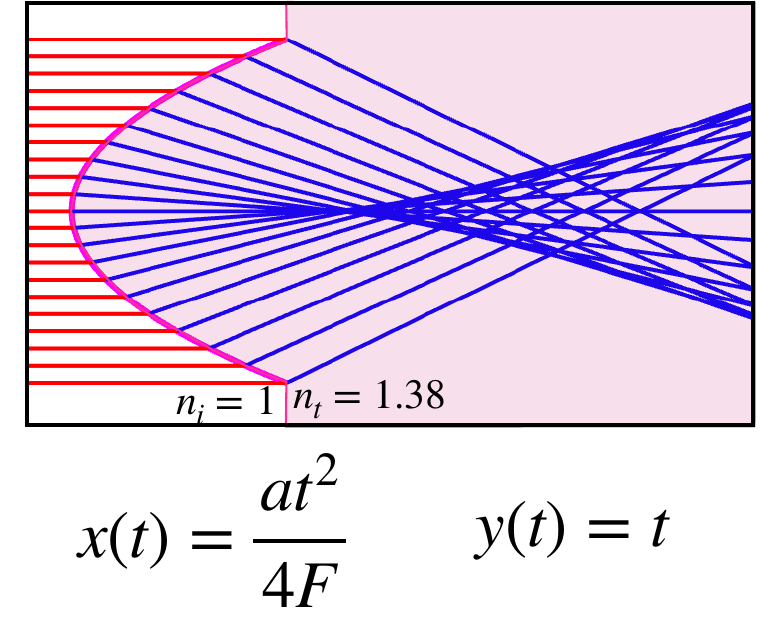}
\caption{Rays refracted by a parabolic surface. $F$ is the focus of the parabola.}
\label{Fig:ParabolicT}
\end{figure}

\begin{figure}[h!]
\centering
\includegraphics[width=0.65\linewidth]{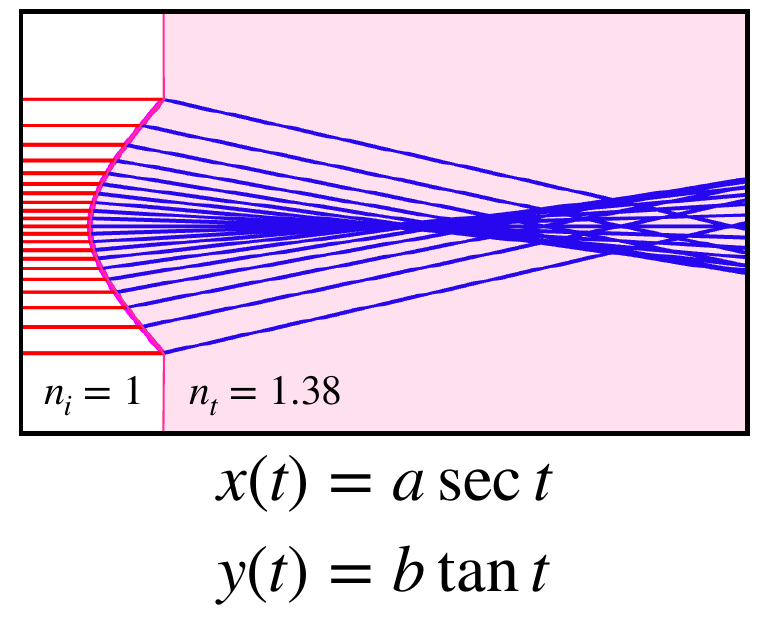}
\caption{Rays refracted by a hyperbolic surface.}
\label{Fig:HyperbolicT}
\end{figure}

We can also calculate the transmitted rays through a circular, elliptical, parabolic, or hyperbolic surface, as shown in Figs. \ref{Fig:CircularT}, \ref{Fig:ElipticoProlateT}, \ref{Fig:ElipticoOblateT}, \ref{Fig:ParabolicT}, and \ref{Fig:HyperbolicT}, respectively.

\section{Total Internal Reflection}

So far, all the examples we have performed meet the condition $n_{i}>n_{t}$. This is because the transmitted angle, $\theta_{t}$, given by Eq. (\ref{Eq:ThetaT}), always results in a real value under these conditions. However, if we consider the case where $n_{t}>n_{i}$, there will be an angle of incidence, $\theta_{i}$, beyond which $\theta_{t}$ becomes imaginary \cite{Lee}. This angle is known as the critical angle and is calculated using the following formula:
\begin{equation}
\theta_c = \sin^{-1}\left(\frac{n_{t}}{n_{i}}\right).
\end{equation}

\begin{figure}[h!]
\centering
\includegraphics[width=\linewidth]{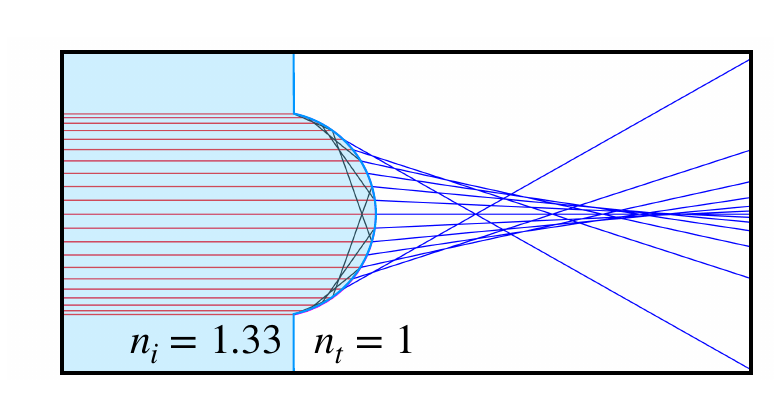}
\caption{Representation of total internal reflection through a circular surface with $n_{i}>n_{t}$.}
\label{Fig:TIR}
\end{figure}

Physically, this equation tells us that when a ray passes from a medium with a higher refractive index (denser medium) to a medium with a lower refractive index (less dense medium), the following phenomena occur:
\begin{enumerate}
\item If the angle of incidence is less than the critical angle, the ray refracts out of the denser medium.
\item If the angle of incidence is equal to the critical angle, the refracted ray travels along the boundary.
\item If the angle of incidence is greater than the critical angle, the light ray is completely reflected back into the denser medium. This phenomenon is known as total internal reflection
\end{enumerate}

For example, the critical angle, $\theta_{c}$, for the water-air interface (where $n_{\text{water}}=1.33$ and $n_{\text{air}}=1$) can be calculated as follows:
\begin{equation}
\theta_c=\sin^{-1}\left(\frac{1}{1.33}\right)\approx48.6^\circ
\end{equation}
Thus, any ray hitting the water-air boundary at an angle greater than 48.6° will undergo total internal reflection, as shown in Fig. \ref{Fig:TIR}.

\section{Conclusions}

In this paper, we have presented a general method for implementing Snell's law and the law of reflection on a chosen curve, using only geometry and basic mathematics, such as differential calculus. This approach allows for the analysis and prediction of the behavior of light and sound as they encounter curved surfaces, facilitating the understanding of complex optical and acoustic phenomena.

Furthermore, we have shown a series of practical examples that illustrate how to apply these laws to different types of curves. These examples demonstrate the versatility and usefulness of our method in calculating Snell's law and the law of reflection and its applicability to Freeform Optics with rotational symmetry.

We conclude that this approach is a powerful tool for education and research in both optics and acoustics. With the basic ideas and tools used here an interesting challenge would be to extend the problem to a three dimensional surface using vector calculus.

\bibliography{ref}
\end{document}